\documentclass[12pt]{iopart}
\usepackage{times}
\usepackage{graphicx}

\begin{document}

\title{Anyon exciton revisited}
\author{D G W Parfitt and M E Portnoi}
\address{School of Physics, University of Exeter, Exeter EX4 4QL, UK}
\begin{abstract}
We review the main results of the anyon exciton model in light 
of recent criticism by W\'{o}js and Quinn. We show that the 
appearance of fractionally charged anyon ions at the 
bottom of their numerically calculated excitation spectra is an 
artefact caused by finite-size effects in a spherical geometry.
\end{abstract}

\section{Review of anyon exciton model}

The anyon exciton model \cite{RP93,PR96} provides a full 
classification of the multiple-branch spectra of a 
four-particle anyon exciton, a neutral composite particle 
consisting of a valence hole and three anyons with charge $-e/3$.
This model is only applicable for large enough separation between 
a photoexcited hole and a two-dimensional electron gas at 
exact fractional filling factor, when the Coulomb field from 
the hole cannot destroy the incompressible quantum liquid.

\begin{figure}
\begin{center}
\includegraphics[width=13.5cm,keepaspectratio]{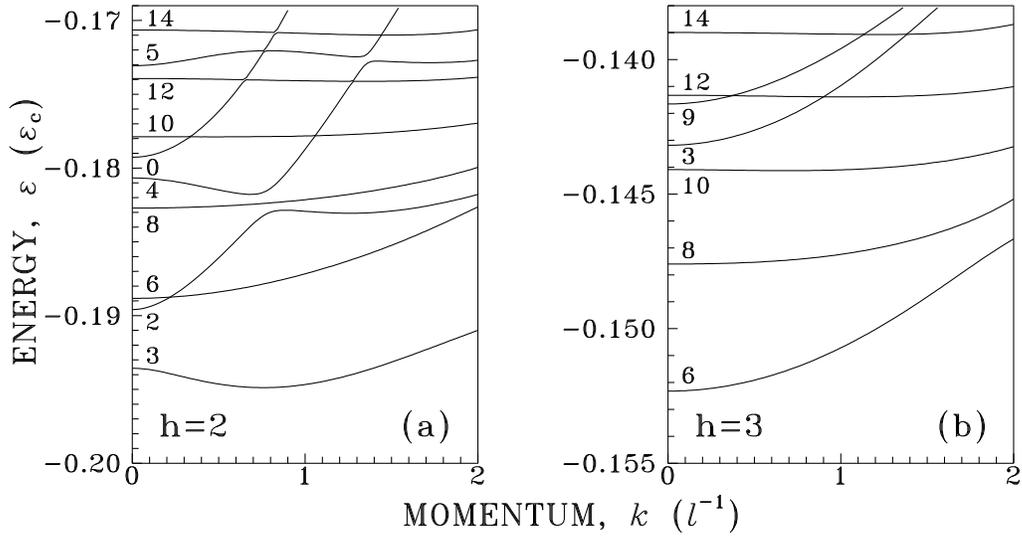}
\end{center}
\caption{Anyon exciton dispersion $\epsilon (k)$ for two values
of electron-hole separation $h$. Numbers show $L$ values;
$h$ is in units of magnetic length $l$.}
\label{Fig1}
\end{figure}

\begin{figure}
\begin{center}
\includegraphics[width=13.5cm,keepaspectratio]{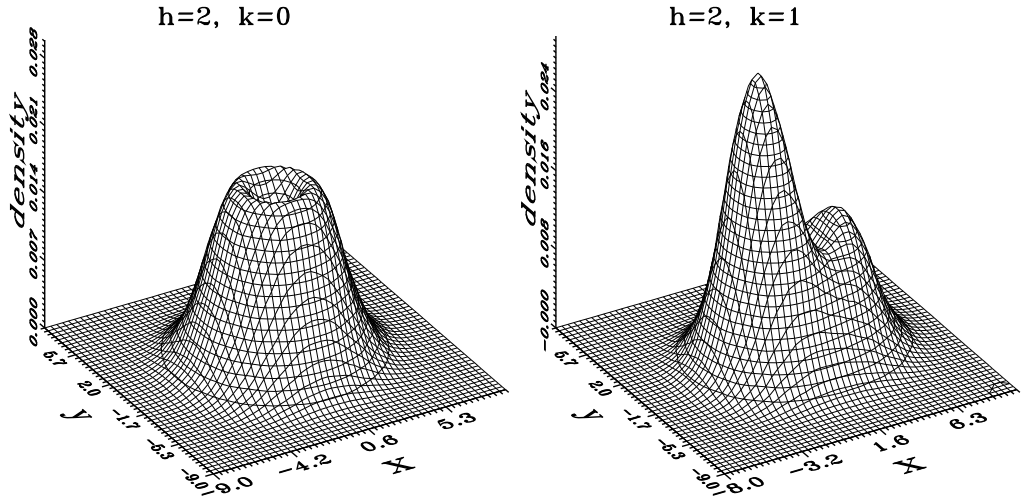}
\end{center}
\caption{Electron density distribution in an anyon exciton for different 
values of the exciton in-plane momentum $k$. The distance $h$ between the 
hole and the incompressible electron liquid is equal to two magnetic lengths. 
The hole is at the origin; the $x$-axis is chosen in the direction of the 
exciton dipole moment.}
\label{Fig2}
\end{figure}

As a neutral particle, the anyon exciton possesses an
in-plane momentum $\mathbf{k}$. At $k=0$ it is also
described by the angular momentum $L_z=-L$, where $L$ is the
degree of the polynomial symmetric with respect to interchange
of anyon coordinates. There is an additional quantum number
which enumerates different polynomials of the same degree.
An important parameter of the problem is the separation $h$
between electron and hole confinement planes, measured
in units of magnetic length $l$.
Two examples of numerically calculated energy spectra
for $h=2$ and $3$ are shown in Fig.~\ref{Fig1}.
For $h=2$, the minimum of the spectrum occurs for non-zero
$k$, which means that the exciton in the lowest
energy state possesses a non-zero dipole moment.
The negative dispersion in Fig.~1a arises because of the
mutual repulsion of $L=2$ and $L=3$ branches.
For higher values of $h$, the ground state is always at $k=0$,
as in Fig.~1b, and the lowest-branch dispersion is always
positive.
The value of $L$ at $k=0$ for the lowest branch increases
with increasing $h$, and thus direct optical transitions from
the ground state are forbidden.
However, for $k\neq 0$ the wavefunction is a mixture of states
with different $L$ values, and magnetoroton-assisted transitions
become possible.
Fig.~\ref{Fig2}, which shows the negative charge distribution around
a valence hole for two different values of $k$, implies a simple 
qualitative picture of such a transition. 
A quasihole appears as a result of recombination of the hole with 
two negatively-charged anyons which are close to it.
The quasihole and a split-off anyon form a magnetoroton.
Notably, for the particular value of $h=2$ used in Figs.~1a and 2,
the asymmetric $k\neq 0$ distribution has a lower energy than the
symmetric one.
Nevertheless, we wish to emphasise that the exciton remains 
neutral, contrary to the statements in Ref.~\cite{WQ}.

Such features of the anyon exciton model as multiple-branch spectra,
dark ground states, and absence of $L=1$ states coincide with the
results of exact diagonalisation for a few-electron system in the
spherical geometry \cite{WQ,AR}.
Similarities between the electron density distribution and pair 
correlation functions for low-lying states in planar and spherical 
geometries are discussed in Ref.~\cite{PR96}.
We wish to emphasise that such a comparison is only valid in the
region of intermediate electron-hole separation, $h\approx 2$,
where both approaches are applicable.

The presence of a `tight' $(L=0)$ state in the upper part of the
exciton energy spectrum, together with a magnetoroton-assisted
transition from the lower part of the spectrum, may explain the
double-line feature observed in the intrinsic spectroscopy of
incompressible quantum liquids \cite{Plentz}.

\section{Classical limit in planar and spherical geometry}

When the separation $h$ exceeds two magnetic lengths 
(which is a requirement for the anyon exciton model to
be valid), the ground state at $k=0$ is formed
by the states with angular momentum obeying the superselection
rule $L=3N$, where $N\geq 2$. 
All these states are hard-core states, i.e. their wavefunctions
are zero if any two of the anyon coordinates coincide. 
This means that the three anyons in the exciton form an equilateral 
triangle; this is in complete agreement with the classical picture.
In fact, for large values of separation, the result of a 
quantum mechanical calculation for the ground state energy 
is the same as that obtained by a simple classical approach.
This is true for quasielectrons obeying both anyonic and bosonic
statistics \cite{PR96}.

\begin{figure}
\begin{center}
\includegraphics[width=13.5cm,keepaspectratio]{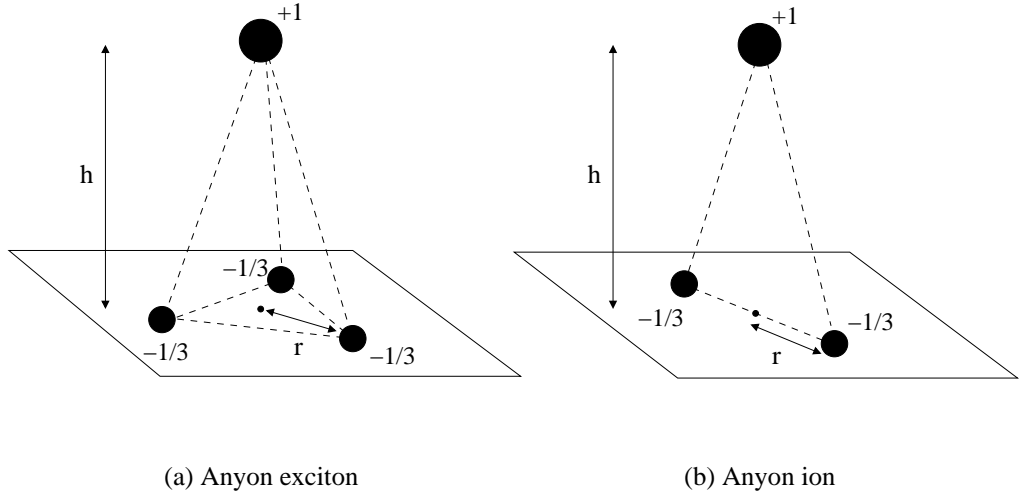}
\end{center}
\caption{`Classical' anyon exciton (a) and ion (b) in the planar geometry,
where $h$ is the separation between the electron and hole confinement planes 
and $r$ is the distance between each anyon and the centre of negative charge.
Note that charges are scaled by the electron charge $e$.}
\label{Fig3}
\end{figure}

\begin{figure}
\begin{center}
\includegraphics[width=13.5cm,keepaspectratio]{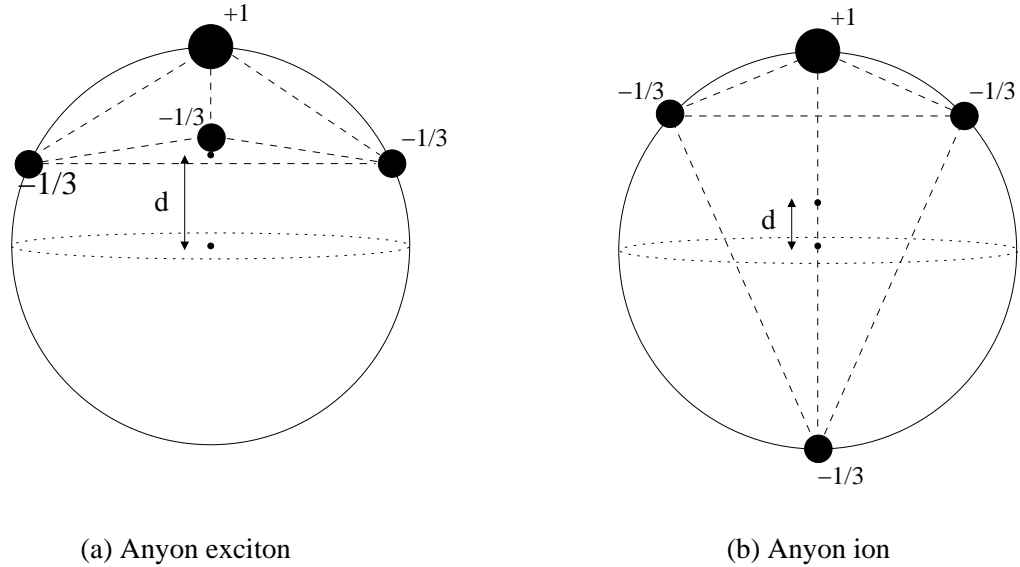}
\end{center}
\caption{`Classical' anyon exciton (a) and ion (b) in the spherical 
geometry, where $d$ is the distance between the centre of negative charge
and the centre of the sphere.}
\label{Fig4}
\end{figure}

It is natural to compare the results of classical calculations
for anyon excitons and ions in both planar (Fig.~\ref{Fig3})
and spherical (Fig.~\ref{Fig4}) geometries.
These calculations are based on minimising the classical potential 
energy of the system, and the hole and anyons are considered
as point charges.
Fig.~3a shows the `classical' $k=0$ anyon exciton, the energy of
which is given by  
\begin{equation}
E=\frac{1}{3\sqrt{3}r}
-\frac{1}{{\left(h^2+r^2\right)}^{1/2}}\, ,
\end{equation}
where $r$ is the distance from each anyon to the centre of
negative charge. Here, and in what follows, charge is
measured in units of electron charge $e$.
Minimising this energy with respect to $r$ gives
\begin{equation}
E_{min}=-\left(\frac{2}{3}\right)^{3/2}\frac{1}{h} \approx -\frac{0.544}{h}\, .
\end{equation}
This result coincides exactly with quantum mechanical calculations 
in the limit of large $h$ \cite{PR96}. A similar calculation for an ion
(Fig.~3b) gives a potential energy
\begin{equation}
E=\frac{1}{18r}-\frac{2}{3{\left(h^2+r^2\right)}^{1/2}}\, ,
\end{equation}
which, when minimised, leads to
\begin{equation}
E_{min}=-\frac{(12^{2/3}-1)^{3/2}}{18h}\approx -\frac{0.485}{h}\, .
\end{equation}
Therefore, as expected, the neutral exciton is more energetically
favourable than a positively-charged ion.

For a spherical geometry, with the valence hole at the north pole
of the sphere, the anyon exciton corresponds to the configuration
shown in Fig.~4a, whereas for the ion, one of the anyons goes to
the south pole of the sphere (Fig.~4b), which corresponds to
infinity on the plane.
Note that the electron-hole separation is taken into account by
introducing an effective interaction $V=-(r_c^2+h^2)^{-1/2}/3$, 
where $r_c$ is taken as the anyon-hole chord length. 
The expression for the energy (with the radius of the sphere
taken as unity) is
\begin{equation}
E=\frac{\sqrt{3}}{9{\left(1-d^2\right)}^{1/2}}
-\frac{1}{{\left(h^2-2d+2\right)}^{1/2}}\, ,
\end{equation}
where $d$ is the distance between the centre of negative charge
and the centre of the sphere.
Minimising this expression in the limit of large electron-hole
separation $h$, we obtain
\begin{equation}\label{eq1}
E_{min}=\frac{1}{3\sqrt{3}}-\frac{1}{h}+\frac{1}{h^3}
-\frac{3}{2h^5}+O\left(\frac{1}{h^6}\right)\, .
\end{equation}
A similar calculation for the configuration in Fig.~4b gives
a potential energy
\begin{eqnarray}
E&=&\frac{\sqrt{3}}{27{\left(1+d\right)}^{1/2}{\left(1-3d\right)}^{1/2}}
+\frac{2\sqrt{3}}{27{\left(1+d\right)}^{1/2}} \nonumber \\
&&-\frac{2}{3{\left(h^2-3d+1\right)}^{1/2}}
-\frac{1}{3{\left(h^2+4\right)}^{1/2}}\, ,
\end{eqnarray}
which, when minimised, yields
\begin{equation}\label{eq2}
E_{min}=\frac{1}{3\sqrt{3}}-\frac{1}{h}+\frac{1}{h^3}
-\frac{9}{4h^5}+O\left(\frac{1}{h^6}\right)\, .
\end{equation}
Comparing Eqs.~\ref{eq1} and \ref{eq2}, one can see that for
large electron-hole separation in a spherical geometry,
the positively-charged ion is energetically more favourable 
than the neutral exciton, which is not the case in the more 
realistic planar geometry.

Introducing realistic form factors \cite{WQ}, which reduce
anyon-anyon repulsion at large distances, is unlikely to
push one of the anyons to infinity and make the ion
energetically favourable in the planar geometry.

In conclusion, we believe that the appearance of 
fractionally-charged anyon ions at the bottom of the numerically 
calculated excitation spectra \cite{WQ} is an artefact caused by 
finite-size effects in the spherical geometry.


\section*{References}
\begin{thebibliography}{9}
\bibitem{RP93}
Rashba E I and Portnoi M E 1993 Phys. Rev. Lett. 70 3315-3318
\bibitem{PR96}
Portnoi M E and Rashba E I 1996 Phys. Rev. B 54 13791-13806 
\bibitem{WQ}
W\'{o}js A and Quinn J J 2001 Phys. Rev. B 63 0453031-04530313; 
{\it ibid.} 63 0453041-04530411; Solid State Commun. 118 225-229
\bibitem{AR}
Apalkov V M and Rashba E I 1995 Solid State Commun. 95 421-424
\bibitem{Plentz}
Plentz F, Heiman D, Pinczuk A, Pfeiffer L N, and West K W 1996
Surf. Sci. 361/362 30-33, and references therein
\end{thebibliography}
\end{document}